# Microring Resonators Coupling Tunability by Heterogeneous 2D Material Integration

R. Maiti[1], R. Hemnani[1], R. Amin[1], Z. Ma[1], M.H. Tahersima[1], T.A. Empante[2], H. Dalir[3], R. Agarwal[4], L. Bartels[2] and V. J. Sorger[1]*

[1]Department of Electrical and Computer Engineering, George Washington University, Washington, DC 20052, USA

[2]Chemistry and Materials Science and Engineering, University of California, Riverside, California 92521, USA

[3]Omega Optics, Inc. 8500 Shoal Creek Blvd., Bldg. 4, Suite 200, Austin, TX 78757, USA

[4]Department of Materials Science and Engineering, University of Pennsylvania, Philadelphia, PA 19104, USA

*Corresponding Author E-mail: *sorger@gwu.edu*

**Abstract**

Atomically thin 2D materials provide a wide range of basic building blocks with unique properties, making them ideal for heterogeneous integration with a mature chip platform. An understanding of the role of excitons in transition metal dichalcogenides in Silicon photonic platform is a prerequisite for advances in optical communication technology, signal processing, and possibly computing. Here we demonstrate passive tunable coupling by integrating few layers of MoTe$_2$ on a micro-ring resonator. We find a TMD-to-ring's circumference coverage length ratio to place the ring into critical coupling to be about 10% as determined from the variation of spectral resonance visibility and loss as a function of TMD coverage. Using this TMD–ring heterostructure, we further demonstrate a semi-empirical method to determine the index of a 2D material ($n_{MoTe2}$ of 4.36+0.011$i$) near telecommunication-relevant wavelength.

# INTRODUCTION:

Silicon photonics is poised to revolutionize a number of diverse fields including data centers, high-performance computing, and photonic information processing [1,2] due to its high index contrast and the dense integration of photonic circuits over a large scale using matured CMOS fabrication technology [3]. However, the rapid increase of internet uses from web applications, cloud computing, and internet-of-things demand miniaturized communication systems to squeeze large numbers of discrete photonic components on a single platform to achieve higher bandwidths [4]. The implementation of an optical switch can aid this process by achieving high bandwidth, low latency, and controllable switching mechanism [5-8]. Micro-ring resonators (MRR) are one such key components of silicon photonics that exhibit capability in various all-optical applications such as optical filter, lasing, optical delay line, and switching [9-11].

However, it is still an open challenge to find a single material able to provide electro-optic functionality, gain, low-loss, and strong light-matter-interactions in integrated photonics. Alternatively, hybrid materials and heterogeneous integration solutions are currently being explored, particularly on silicon substrates for compatibility with microelectronics. For instance, the integration of germanium as a detector and III–V compound semiconductors for light sources are technologically challenging on a silicon substrate due to mismatched lattice constants and thermal expansion coefficients, despite some efforts having been made [12]. Other active opto-electronic materials such as transparent conductive oxides, while showing high switching performance, usually introduce relatively high optical losses [13]. Hence, heterogeneous integration of two-dimensional (2D) transition metal dichalcogenides (TMDs) is essential to realize a hybrid and compact photonic integrated circuit since they provide a highly tunable and ultra-thin platform with absence of (or reduced) covalent bonding [14,15].

2D materials are actively investigated due to their interesting properties in various fields of physics, chemistry, and materials science, starting with the isolation of graphene from graphite [16]. While graphene shows many exceptional properties, its lack of an electronic bandgap has stimulated the search for alternative 2D materials with semiconducting characteristics [17-20].

TMDs, which are semiconductors of the type $MX_2$, where M is a transition metal atom (such as Mo or W) and X is a chalcogen atom (such as S, Se or Te), provide a promising alternative [21]. 2D TMDs exhibit unique physical properties such as indirect to direct bandgap transition [22], quantum confinement [23], strong spin-orbit coupling [24], and high exciton binding energies [25] as compared to their bulk counterparts. This makes them highly promising for both new fundamental physical phenomena as well as innovative device platforms ranging from photo detection [26,27], modulation [28], light-emitting diodes, [29], and lasers [30]. Moreover, optical response of the 2D materials can be modulated by tuning their carrier density via electrical or optical means that modify their physical properties (e.g., Fermi level or nonlinear absorption), making them versatile building blocks for optical modulators or reconfiguration in a Si-based hybrid platform.

Here we demonstrate heterogeneous integration of few layers of $MoTe_2$ on Si photonics platform for the first time. The interaction between the TMD and the Si MRR provides novel tunable coupling phenomena that can be tuned from over coupling to under coupling regime via critical coupling condition by means of altering the rings effective index via the integration of TMD. We analyze the coupling physics and extract fundamental parameters such as quality factor, visibility, transmission at resonance etc., as a function of TMD coverage on the ring. We find a critical-coupling coverage-ratio value (~10%) for a given ring resonator which is relevant for device functionality. Furthermore, we determine the index ($n_{MoTe2} = 4.36+0.011i$) of the few layered TMD at 1.55 μm wavelength in a semi-empirical approach using the ring resonator as an index sensor platform.

**RESULTS & DISCUSSION**:

In order to obtain an understanding of phase modulation in heterogeneous integrated systems it is important to understand the interaction between TMDs ($MoTe_2$, $MoS_2$, $WS_2$ etc.) and a MRR. Our system to analyze such phase modulation uses an all-pass type silicon MRR incorporating a few layers of $MoTe_2$ transferred onto a part of the ring (Figure 1a & b). The aim is to improve the waveguide bus-to-ring coupling efficiency by shifting the phase through introducing a few-layered of $MoTe_2$ flakes on top of the resonator. By varying the $MoTe_2$ coverage length we observe a tunable coupling response. To achieve TMD-loading of the rings, we utilize a precise and cross-contamination free transfer using the 2D printer method recently developed [31] (refer to SI1). The transmission spectra before and after TMD transfer reveal definitive improvement of coupling in terms of visibility defined as the transmission amplitude ratio ($T_{max}/T_{min}$) upon TMDs loading. This hybrid device shifts towards the critical coupling regime as compared to before transfer where it was over coupled.

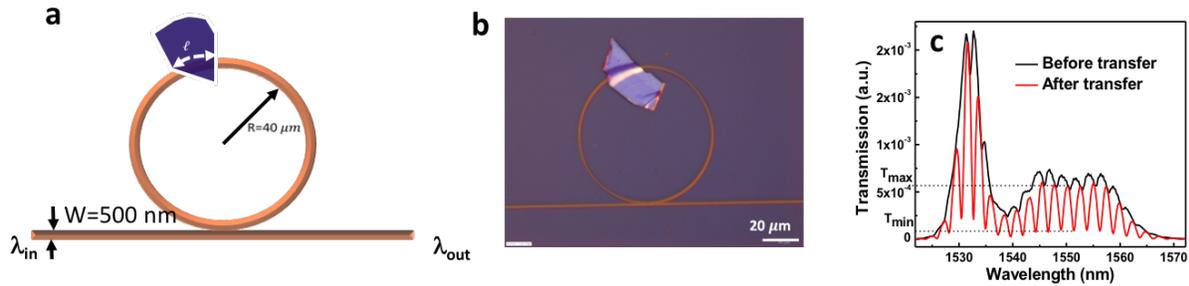

**Fig. 1. TMD loaded micro-ring resonator (MRR)** (a) schematic and (b) optical microscope image of a MRR (R= 40 μm & W= 500 nm) covered by a MoTe$_2$ flake of length (*l*) precisely transferred using our developed 2D printer technique. (c) Transmission output before and after the transfer of MoTe$_2$ showing improvement of coupling efficiency as it brings the device close to critically coupled regime after the transfer of the TMDs layer.

We fabricated a set of ring resonators with different percentages of MoTe$_2$ coverage between 0% and 30% (Figure 2) which allowed us to extract different parameters to understand coupling physics. The cavity quality (*Q*) factor is found to decrease from 1600 to 900 as the ring coverage is raised from 0 to 27% (figure 2a). We attribute this as gradual increase of loss arising from both MoTe$_2$ absorption near its band edge corresponding to indirect bandgap of 0.88 eV [32], and the small impedance mismatch between bare and TMD-covered sections of the rings. In contrast to the monotonic behavior of the quality factor, the minimum transmission ($T_{min}$) at resonant wavelength initially decreases until 10% coverage is reached and then increases for higher coverages (figure 2b). The visibility ($T_{max}/T_{min}$) shows the opposite trend being maximal near 10% of TMD coverage (figure 2c). In combination, these findings indicate tunability of the coupling condition by means of varying TMD coverage.

The performance of a ring resonator is determined by two coefficients: the self-coupling coefficient (r), which specifies the fraction of the light transmitted on each pass through the coupler; and the round-trip transmission coefficient (a), which specifies the fraction of the light transmitted per pass around the ring. For the critical coupling condition, i.e. when the coupled power is equal to the power loss in the ring $1-a^2=k^2$ or $a=r$, ($k$=cross coupling coefficient), the transmission at resonance becomes zero. At this point, the difference (|a-r|) is found to be minimum at ~10% of coverage (figure 2d) suggesting close to critical coverage since |a-r| is inversely proportional to the square root of visibility term [33]. Thus, the coupling condition is tunable from the over coupled regime (r<a) to under coupled (r>a) via the critically coupled regime (r=a) as a function of TMD coverage. Being able to operate at critical coupling is important for active device functionality; for instance, the extinction ratio of an MRR-based electrooptic modulator is maximized at critical coupling [34,35].

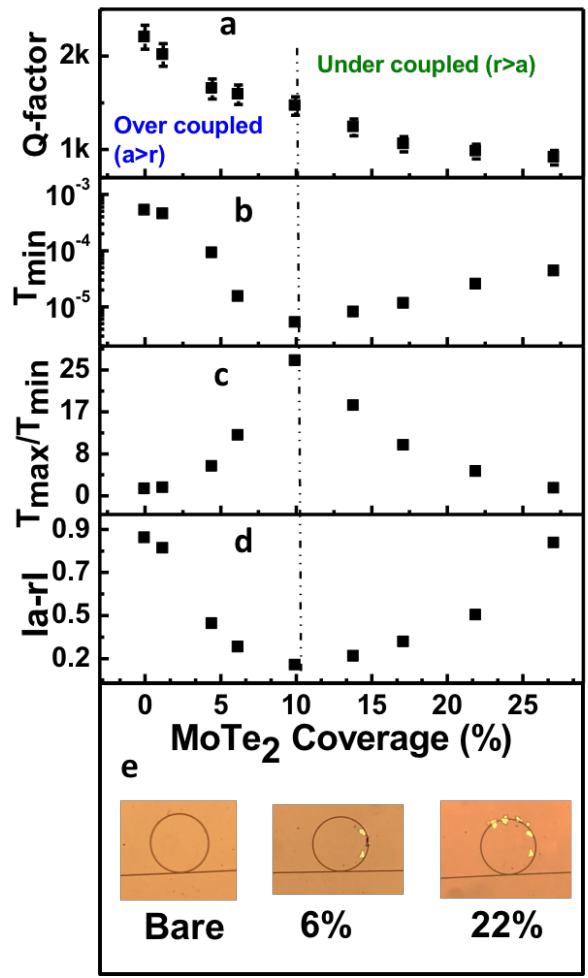

**Fig. 2. TMD loaded tunable coupling effect**. Variation of (a) *Q*-factor; (b) Minimum transmission, $T_{min}$; (c) Visibility ($T_{max}/T_{min}$); (d) Difference between the self-coupling coefficient and round-trip transmission coefficient, |*a-r*|; as a function of MoTe$_2$ coverage. The monotonic decrease of *Q*-factor suggests increasing loss for higher coverage, i.e. more transferred TMD. Tunability of coupling effect i.e. transition from over-coupled to under-coupled regime via critically-coupled condition (dashed vertical line) is evident from the variation of $T_{min}$, $T_{max}/T_{min}$ & |*a-r*|. (e) The corresponding figures for different coverages (bare ring, 6% & 22%) showing the advantages of precise transfer by 2D transfer methods.

In order to understand, the coupling mechanism of a ring resonator, it is important to extract and distinguish coupling coefficients (*a* and *r*), as they are governed by different factors in design and fabrication. However, it is not possible to decouple both the coefficients without additional information, since *a* and *r* can be interchanged (eqn 1). The transmission from an all-pass MRR (figure 3a) is given by

$$T_n = \frac{a^2 + r^2 - 2ar\cos\varphi}{1 + r^2 a^2 - 2ar\cos\varphi} \quad (1)$$

where, $\varphi$ is the round-trip phase shift, *a* is round-trip transmission coefficient related to the power attenuation coefficients by,

$$a^2 = \exp(-\alpha_{Si}(2\pi R - l)) * \exp(-\alpha_{TMD-Si} * l) \quad (2)$$

where *l* = TMD coverage length, $\alpha_{Si}$ and $\alpha_{TMD-Si}$ are the linear propagation losses for Si waveguide and TMD-transferred portion of the Si waveguide in the ring, respectively. We find the propagation loss for Si and TMD-Si to be 0.008 dB/μm (Figure SI3) and 0.4 dB/μm (Figure 3b), respectively via the cutback method at 1550 nm.

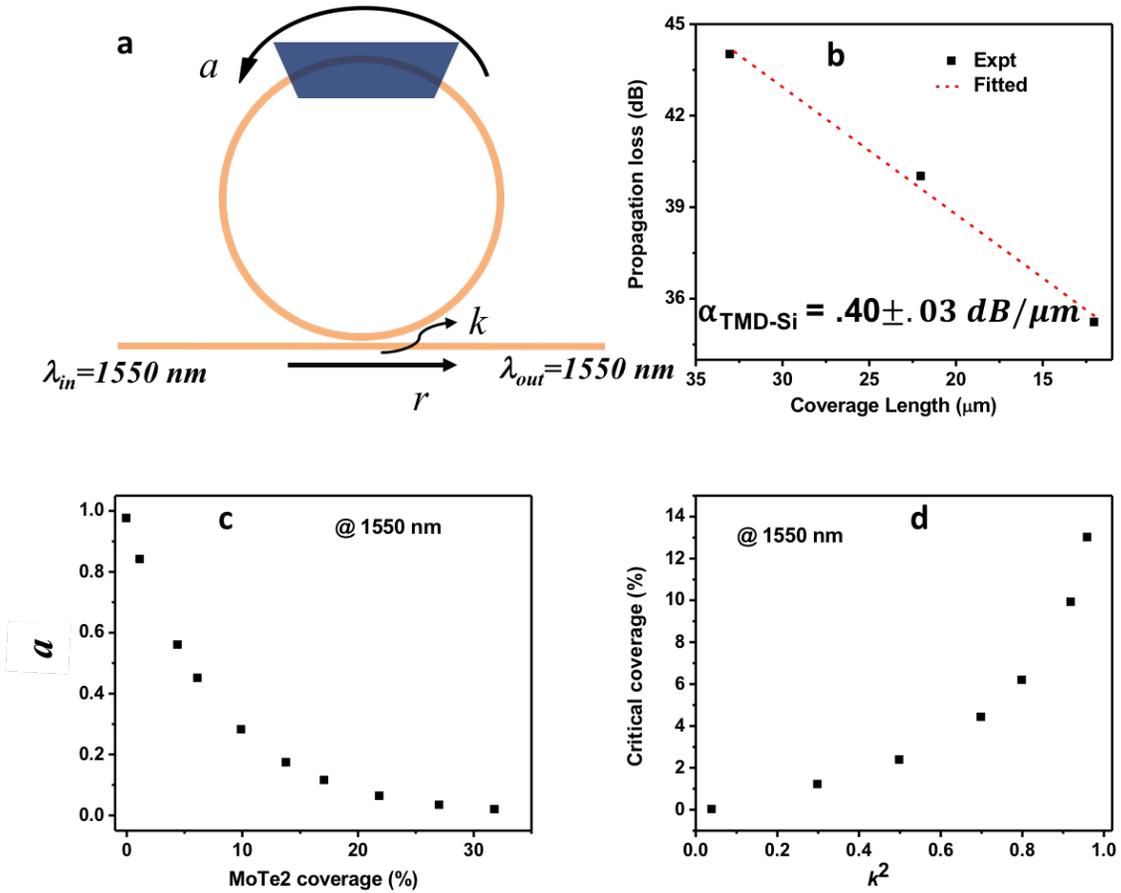

**Fig. 3. Coupling coefficients for TMD integrated hybrid Si MRR** a) Schematic representation of MRR showing self-coupling coefficient ($r$), cross-coupling coefficient ($k$) and round-trip transmission coefficient ($a$) where the ring is partially covered by a MoTe$_2$ flake. (b) The propagation loss ($\alpha_{TMD\text{-}Si}$) for TMD-covered portion of the ring is found to be 0.4 dB/µm using cutback measurement. Tunability of (c) round-trip transmission coefficients explains the exponential decrease as a function of flake coverage. (d) Relationship between critical coverage (%) and power coupling co-efficient ($k^2$) assuming lossless coupling ($r^2+k^2=1$).

Inserting these values into (2), we find the round-trip transmission coefficients ($a$) to be tuned as a function of TMD coverage (figure 3c). The variation of $a$ from 0.97 to 0.01 as a function of coverage confirms the transition from over-coupled to under-coupled regime since $a = 1$ suggests that there is no loss in the ring. Hence, the loss tunability in MRRs can be manipulated accordingly by controlling the coverage length [36]. The MRR transmission at resonance leads to the form, $T_{n,res} = \left(\frac{a-r}{1-ar}\right)^2$, therefore, the critical-coverage ($a=r$) anticipates zero-output transmittance. Hence, for a given MRR (fixed $k^2$), the critical coverage value could be determined provided lossless coupling ($r^2+k^2=1$) (figure 3d).

Si-based MRRs provide a compact and ultra-sensitive platform as refractive index sensor to find an unknown index of the 2D materials at telecom wavelength (1.55μm). In essence, the shift in resonant wavelength can be used to sense the optical properties effecting entities on the silicon core or the cladding [37]. We observe a gradual resonance red-shift from bare to increased coverage of MoTe$_2$ of 4.5%, 10% and 17%, respectively (figure 4a). The resonant wavelength ($\lambda_{res}$) of a MRR is proportional to the effective refractive index of the propagating mode in the circular waveguide [38]. Therefore, the change in effective mode index ($\Delta n_{eff}$) is related to change in resonance ($\Delta\lambda$) following $\Delta n_{eff} = \frac{\Delta\lambda}{\lambda_{res}} * n_{eff,control}$, where, $n_{eff,control}$ is the effective mode index for Si MRR (i.e. without any TMD flakes transferred). The effective index for the control sample (Si ring+SiO$_2$ cladding) can be found from FEM Eigenmode analysis choosing the TM-like mode in correspondence with our TM-grating designs used in measurements (figure SI4). We map out the resonance shift ($\Delta\lambda$) as a function of MoTe$_2$ coverage (figure 4b, (i)) for needed calibration to determine the unknown index using a semi-empirical approach. The positive change in the effective index ($\Delta n_{eff}$) as a function of MoTe$_2$ coverage relates to an increased effective mode index with TMD transfer (Fig. 4b), (ii)) and corresponding red-shifts thereof.

Now, we obtain a resonance shift of 1 nm for 4 % coverage, which gives a corresponding change in effective index of 0.001 (figure 4b). At this point, it is important to establish the relation between the effective mode index with the refractive index of the unknown TMDs. Since, in our case, the MRR is partially covered by MoTe$_2$ flakes and the resonance shift arising from change in effective mode index due to the change in coverage length, the effective refractive index of the ring can be formulated as an effective length-fraction index via

$$n_{eff,ring} = \frac{(2\pi R - l) * n_{eff,control} + l * n_{eff}}{2\pi R} \qquad (3)$$

where, $R$ is the radius of the ring and $l$ is the MoTe$_2$ coverage length. Using (3), it is straightforward to find $n_{eff}$, after TMD transfer.

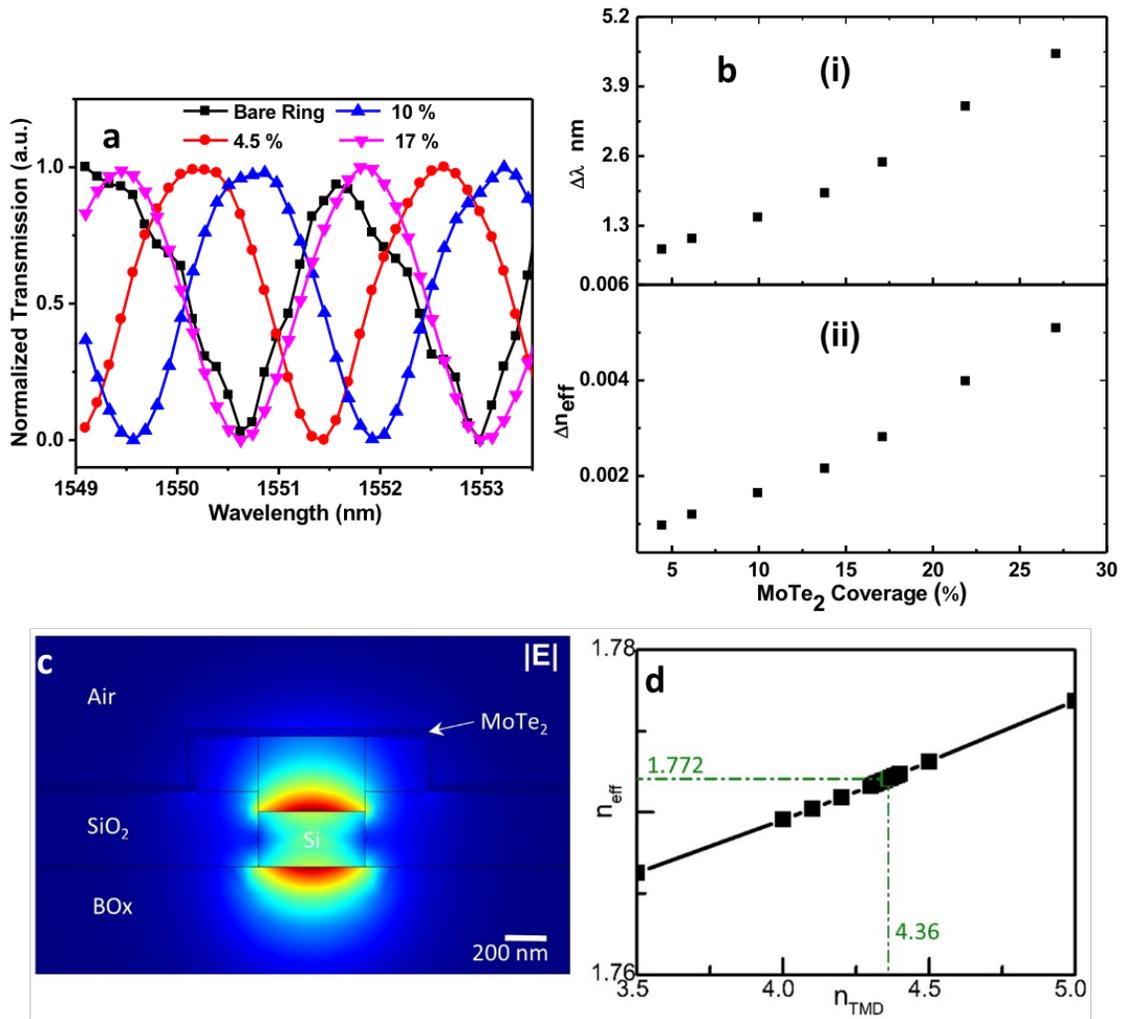

**Fig.4. Ring resonator as refractive index sensor** (a) Normalized transmission spectra for different coverages showing gradual red shift. Variation of (b) Resonance shift (Δλ) and effective index change ($\Delta n_{eff}$) extracted from (a) as a function of MoTe$_2$ coverage length. The mapped-out resonance shift as a function of coverage length provide the calibration curve to determine the unknown index of the TMDs. (c) Mode profile (|E|) for the portion of the ring with MoTe$_2$ transferred flakes from Eigenmode analysis, and (d) FEM results for effective index, $n_{eff}$ vs. the MoTe$_2$ index, $n_{TMD}$ to extract the material index from experimental results. The green dashed line exhibits the obtained MoTe$_2$ index from our results as 4.36.

Once, $n_{eff}$ is known, we can find the effective index of the heterogeneous optical mode (SOI plus TMD) through Eigenmode analysis for a device geometry and provide experimentally measured thickness values of the flake (figure SI5). By sweeping the value of the TMD ($n_{TMD}$ range = [3.5, 5]) material index in cross-sectional Eigenmode analysis of the waveguide structure in the MoTe$_2$ transferred section of the ring (figure 4c), we can match the numerically obtained effective index with that found from aforementioned experimental results. We find the index of the bulk MoTe$_2$ material to be 4.36 (figure 4d). This index value is closely aligned with reported values in previous studies for bulk MoTe$_2$ [38]. We also find the imaginary part of the material index from the cutback method varying flake sizes to be ~0.011. This semi-empirical approach has several advantages over conventional ellipsometric technique to determine the refractive index of the unknown materials; for instance, in ellipsometry, large uniform flakes (few mm to cm) are needed since the beam spot size is generally large, which are still challenging to obtain with uniform properties across such scales. However, for the method presented here, small TMD samples (i.e. flakes) can be measured (~500 nm). In fact, the limit of this technique is not bounded by the MRR waveguide width, since partial coverage of the top waveguide is also possible to measure, but with linear phase shift scaling with respect to area covered. Besides this, to determine refractive index of the materials from the ellipsometry data, the experimental data needs to be fitted by an exact physical model which bears ambiguity and thus introduces additional inaccuracy.

**CONCLUSION:**

We have studied the interaction between few layers of MoTe$_2$ and Si MRR for the first time. We observed tunability of the coupling strength i.e. the ring resonator can be tuned from the over-coupled to the under-coupled regime while passing through the critically-coupled point. The underlying physical mechanism of tunable coupling can be explained by extracting different coupling coefficients as a function of coverage length. For the materials properties examined here, the critical coverage value for a given MRR is ~10%. We further demonstrated a semi-empirical approach to determine the index of miniscule (~500 nm) TMD material flakes using the index sensitivity of the ring resonator. These findings along with the developed methodology for placing MRRs into critical coupling and determining the refractive index of 2D materials could be useful tools in future heterogeneous integrated photonic and optoelectronic devices. This developed technique could also be used to determine the optical refractive index of monolayer 2D materials, which is challenging with traditional techniques due to lateral TMD flake size, and atomic thickness of these materials.

**METHODS:**

The Si waveguide and MMR system was fabricated by using a silicon-on-insulator (SOI) substrate, where the Si device layer is 220 nm and the oxide layer is 2 μm thick. The pattern was defined by using electron-beam lithography with negative photoresist (HSQ). Then a Bosch etching process was performed to etch silicon and the HSQ layer performed as an etching mask during the Bosch process. After etching, the HSQ residue was removed by using HF. Then the precise transfers of TMD materials were performed by using our developed 2D printer method (Figure S1) after deposition of ~300 nm of $SiO_2$ cladding layer on Si MRR by plasma enhanced chemical vapor deposition (VERSALIN PECVD). It provides fast and cross-contamination free transfer of flakes having different lengths and thickness obtained from mechanically exfoliated TMD crystals. The precise transfer of the TMD flakes without troubling neighboring devices is key here. Therefore, the role of the micro stamper is a critical aspect for the transfer since for successful transfer of a single flake, effective contact area ($A_{eff}$) of the micro-stamper must be greater than the flake area ($A_{flake}$).

The experimental setup for measuring the hybrid TMD-Si devices is shown in SI2. Briefly, light from a broadband source (AEDFA-PA-30-B-FA) is injected into the grating coupler optimized for the TM mode propagation in the waveguide. The light output from the MRR is coupled to the output fiber by a similar grating coupler, and detected by the optical spectral analyzer (OSA202).

**Acknowledgement**

V.S. and L.B. are supported by the AFOSR under grant number FA9550-17-1-0377 and NSF Materials Genome Initiative under the award number NSF DMREF 14363300/1455050.

TOC

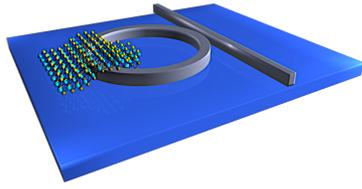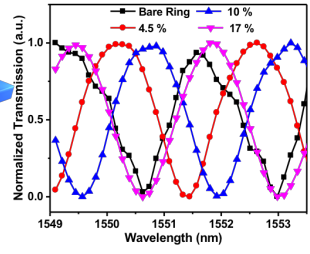